# Wideband THz Low-Scattering Surface Based on Combination of Diffusion and Absorption

Kasra Rouhi, Ali Abdolali[*], and Susan Fallah

*Abstract*— In this paper, a wideband and low-scattering metasurface in terahertz (THz) is introduced. The proposed coding metasurface is composed of four different graphene square patches in one layer, which has a distinct bias voltage. By optimizing the chemical potential of each patch, the reflection phase and amplitude of a designed element can be controlled in a real-time manner. The chemical potential optimizing approach is a promising method to develop metasurfaces, which can tune the reflection phase, magnitude, or polarization dynamically at different frequencies spectrum. Indeed, by adjusting the metasurface reflection profile, the suggested device can manipulate the reflected wave. Also, this metasurface can reduce reflection energy in the wide-band spectrum. The programmable surface disperses reflected power in various directions in a first frequency band and converts incident electromagnetic waves into heat at second frequency band. The obtained results demonstrate that more than 10 dB reflection reduction can be realized over 1.02 to 2.82 THz under both TE and TM polarized wave incidences. Due to the conformal properties of the graphene monolayer, the stealth feature of the metasurface is well preserved while wrapping around a metallic curved object. This optimization method has an excellent aptitude for phase, magnitude, and polarization control in various beamforming applications at the THz spectrum for high-resolution imaging and stealth technology.

## I. INTRODUCTION

Metasurface shows many fascinating features of metamaterials, giving rise to various electromagnetic applications at a subwavelength scale, such as diffusion [1], optical signal processing [2], power divider [3], asymmetric transmission [4], absorber [5], and beam manipulation [6]. Besides, the application of metasurfaces in wave manipulation becomes a hot topic in recent years. By spatially changing the characteristics along the metasurface, one can elaborately engineer structure with desirable functionalities. The Snell's law governs the propagation features of electromagnetic waves at the interfaces of not only natural materials but also artificial metamaterials. The anomalous reflection and refraction, that break the constraint of conventional Snell's law, is one of the most exciting achievements of two-dimensional metasurfaces. Recently, several metasurface-based methods have been introduced to suppress the backscattering energy in two-dimensional structures, such as mantle cloak [7], [8], metamaterial absorber [9], [10], checkerboard metasurface [11], [12], and diffusion metasurface [13]–[15]. Mantle cloak cancels the scattering from planar, cylindrical, or spherical objects. The working principle of absorbers is based on converting incident energy into heat. This method seems to be the right choice for several applications, but the temperature rise is the main drawback of the absorption method in some devices. Recently various wave manipulation applications have been investigated by designing two particular metasurfaces with out of phase reflection properties as the primary coding bits of "0" and "1" and arranging their distribution in the metasurface with a predetermined map [16]. The metasurface with the checkerboard configuration of two anti-phase types of artificial magnetic conductors divides the reflected beams into four analogous lobes. It reflects them toward diagonal and off-diagonal directions. Still, the checkerboard arrangement suffers from fixed scattering directions and high bistatic reflection values. Finally, the diffusion metasurfaces, composed of disorderly distributing anti-phase elements with contrary reflection phases responses, has a promising potential for wave diffusion. As a result, the diffusion metasurface, mimicking a diffuse reflection surface and disperses the scattering waves into numerous directions. Thus, the redirected energy value in a single direction will be significantly reduced in a broad frequency spectrum. In all the methods as mentioned earlier, the design of proper unit-cell for the desired usage is mostly intuitive, achieved by multiple optimization methods, or just chosen by experience.

Different kinds of studies have been employed in designing an optimized structure for field synthesis. Among all of the proposed methods in recent years, topology optimization is a well-known and efficient method. Topology optimization, combined with numerical optimization algorithms, has attracted attention as the computational power of computers increases at a fantastic speed [17]. The concept of topology optimization method was founded in the structural mechanic field several decades ago, whereas this method has been extended to complex structures in various engineering applications [18]. In [19], the authors used a topology optimization approach based on discrete square-shaped pixels. The arrangement of these square metal sub-blocks is optimized by using binary particle swarm optimization to reach the required reflection phase. In other research, Borgese et al. design an ultra-wideband and angularly stable polarization converter [20]. The polarizer device is composed of a periodic composition of miniaturized metallic elements printed on a grounded dielectric substrate. To obtain broadband polarization converting, the metasurface is optimized by a genetic algorithm that imposes the minimization of the amplitude of the co-polar refection coefficient over a broad frequency band. In summary, various applications of topology optimization methods have been investigated in recent years, such as frequency selective surfaces [21], asymmetric transmission [22], wave manipulation [23], and wideband scattering diffusion [24].

Passive metasurfaces, which are composed of nonuniform components with various geometries, can impose changes in characteristics of the incident beam. Nevertheless, passive metasurfaces are unable to reach tunability, as the arrangement of their subwavelength elements is fixed. A wide range of electro-optical materials have recently emerged like graphene [25], indium tin oxide (ITO) [26], and vanadium dioxide [27], [28]. The surface conductivity of graphene is widely tunable by the change of its electrochemical potential by applying an external gate voltage. According to ultrathin thickness, eases of patterning schemes with current technology, and broadband operation, it has been exploited in reconfigurable devices for THz beam manipulation. Graphene-based devices have various applications in secure communication [14], broadband polarizers [29], orbital angular momentum generation [30]. Recently, Torabi et al. employed a novel method of evolutionary optimization for developing the tunable unit-cell for broadband THz frequency selective surfaces [31]. The surface of the suggested structure is divided into 64 pixels, each of which can be gold, biased graphene, doped (unbiased) graphene, or air. Eventually, the Random hill-climbing optimization algorithm was used to find the best scheme for metasurface. A combination of graphene and conventional conductors or different biased graphene enhances the degrees of freedom in the design method.

Here, we propose a new strategy for designing a wide-band and conformal graphene-based coding metasurface to suppress THz incident waves by a combination of diffusion and absorption. Recently, several pieces of researches reported a similar approach for suppressing reflected waves in the GHz band [32], [33], whereas, to the best of the author's knowledge, the real-time manipulation of waves and energy suppression had not been achieved yet via controllable coding metasurfaces in THz spectrum. In this paper, the suggested metasurface designed based on optimizing the chemical potential of graphene patches on the thick substrate. We have defined two different goals for primary optimization. As a first goal, we try to introduce two anti-phase elements in the first frequency spectrum to diffuse incident waves into numerous directions through random distribution. The second goal is absorbing incident power in the adjacent frequency band. The proposed graphene-based metasurface contains four graphene patches, which can be tuned externally. These four pixelated graphene patches are embedded in a similar layer, but each piece has an individual external bias. The suggested metasurface can act as encoded anti-phase elements and absorber simultaneously via employing the proper chemical potential profile in two adjacent frequency bands. So the proposed structure is a versatile device that can manipulate or absorb illuminated THz wave in a real-time manner. The results illustrate that the THz surfaces show low scattering and polarization-insensitive response below –10 dB over a wide frequency band from 1.02 to 2.82. Moreover, this structure keeps its low scattering performance when it is wrapped around a cylindrical metal bar.

## II. Structure design

The primary part of the proposed structure contains graphene as a tunable material. Graphene, a monolayer of carbon atoms arranged in a two-dimensional honeycomb lattice, is a promising material which, due to its fascinating electromagnetic features and the design freedom it offers, has attracted significant attention. This material is sensitive to an external gate bias and can be employed to control the THz waves dynamically. The surface conductivity of graphene can be tuned via external electrostatic bias or external magnetostatic bias by Hall Effect. Also, graphene complex conductivity is a function of frequency and temperature. Because of the graphene single-layer configuration, graphene can be modeled as an exceedingly thin layer characterized by a two-dimensional complex surface conductivity [34]

$$\sigma_s(\omega, \mu_c, \Gamma, T) = \sigma_s^{intra}(\omega, \mu_c, \Gamma, T) + \sigma_s^{inter}(\omega, \mu_c, \Gamma, T) \tag{1}$$

The complex conductivity is a combination of the intraband and interband portions, which $\sigma_{intra}$ and $\sigma_{inter}$ describe intraband and interband parts of complex conductivity. Based on the Pauli-blocking effect, the interband part of conductivity can be neglected at lower THz frequencies and room temperature, so the graphene conductivity central part is an intraband contribution, and each portion can be calculated by Kubo's formula as follows [34]

$$\sigma_s^{intra}(\omega, \mu_c, \Gamma, T) = -j\frac{e^2 k_B T}{\pi \hbar^2 (\omega - j2\Gamma)} * \left[\frac{\mu_c}{k_B T} + 2\ln(e^{-\mu_c/k_B T} + 1)\right] \tag{2}$$

$$\sigma_s^{inter}(\omega, \mu_c, \Gamma, T) = -j\frac{e^2}{4\pi \hbar}\left(\frac{2|\mu_c| - \hbar(\omega - j2\Gamma)}{2|\mu_c| + \hbar(\omega - j2\Gamma)}\right) \tag{3}$$

In which, $K_b$ is the Boltzmann's constant, T is the temperature, $\hbar$ denote the reduced Planck's constant, $\omega$ is the angular frequency, e is the electron charge, $\Gamma=1/2\tau$ is the electron scattering rate, $\tau$ is the transport relaxation time, and $\mu_c$ defines the chemical potential. Chemical potential is a function of external bias and can be tuned dynamically by applying an appropriate external gate voltage. In this paper, the temperature and transport relaxation time are considered constant and equal to T=300 K and $\tau$=0.6 ps, which is a practical value in current manufacturing technologies.

The particular goal of this paper is to develop a reconfigurable digital metasurface in which the operational state of each element can be set independently by external DC voltage. Figure 1 represents the schematic view of the graphene-based digital metasurface, which designed for real-time wave control. This surface is providing real-time control over reflected wavefront at THz frequency. In the first layer, four similar graphene square-shaped patches embedded as a primary element with a distinct DC bias voltage. The periodicity of the subwavelength structure is Lattice_Constant=40 µm. The width of each patch is equal to Patch_Size=17 µm, while these patches are lying on alumina and a doped silicon layer, respectively. These two extra layers inserted to serve electrostatic biasing of graphene patches, whereas the thickness of each layer is ignorable respect to operating wavelength. On the beneath of these auxiliary layers, a grounded quartz thick layer with the depth of Depth=25 µm embedded as a dielectric substrate. The electromagnetic parameter of the quartz substrate is $\varepsilon_r$=3.75 and tan$\delta$=0.0184. Finally, the last layer is gold, which restricts wave transmission through the metasurface.

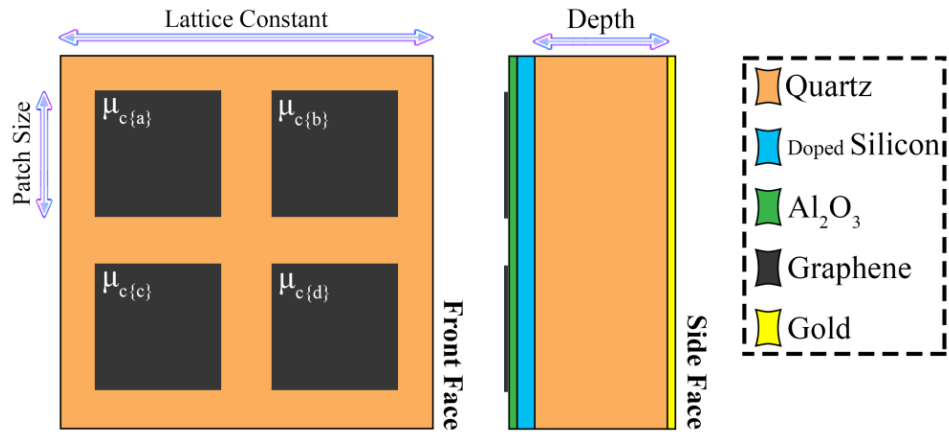

*Figure 1. Schematic view of multifunctional reconfigurable graphene-based unit-cell and constitutive material. By applying various chemical potential profiles, the reprogrammable meta-device can perform reconfigurable wave manipulation or wide-band reflection reduction.*

To realize a 1-bit low reflection digital metasurface, the contribution of two anti-phase coding elements providing "0" and "π" reflection phases is needed. So we utilize CST Microwave Studio to get proper values for chemical potentials to offer two distinct anti-phase metasurfaces in predetermined frequency range considering reflection magnitude near the unit. Each coding metasurface can dynamically possess an arbitrary reflection phase of "0" and "π" to emulate the "0" and "1" phase-encoded bits. The anti-phase response in the predetermined frequency band is our first goal. As a second goal, we tried to decrease the reflection magnitude of each element regardless of the reflection phase in the adjoining frequency band. Attained chemical potential set of {$\mu_{ca}$, $\mu_{cb}$, $\mu_{cc}$, $\mu_{cd}$}={1.3eV/1.1eV/0.55eV/0eV}, and {0eV/0eV/0.3eV/0.85eV} can mimic "0" and "1" states in the first frequency band whereas absorbing incident energy in the second frequency band. All of the required chemical potentials obtained based on optimization by using CST Microwave Studio.

After calculating the required chemical potential for each patch, phase, and amplitude of the 1-bit coding metasurfaces are represented in Figure 2. It is evident that in the first frequency band, each reflection phase has a π difference with an adjacent element, while reflection magnitude in all cases is nearly similar. Also, in the second frequency band, the reflection magnitude is low enough to absorb incident energy while the reflection phases are not equal and have a significant difference. Finally, calculated results confirm that such a unit-cell structure is a qualified candidate for building blocks to realize the wideband reflection reduction. This method is not restricted to meet this specific goal, so by employing another topology or geometry, we can develop structure capability. The graphene patches configuration is an essential part of metasurface design, so selecting an appropriate shape for each segment has a significant influence on design procedure. By using the other complex geometries or optimization methods, it is possible to search for more intricate goals. The previously reported researches have theoretically and experimentally proved the feasibility of such a reconfigurable structure within the realm of current fabrication technology [35].

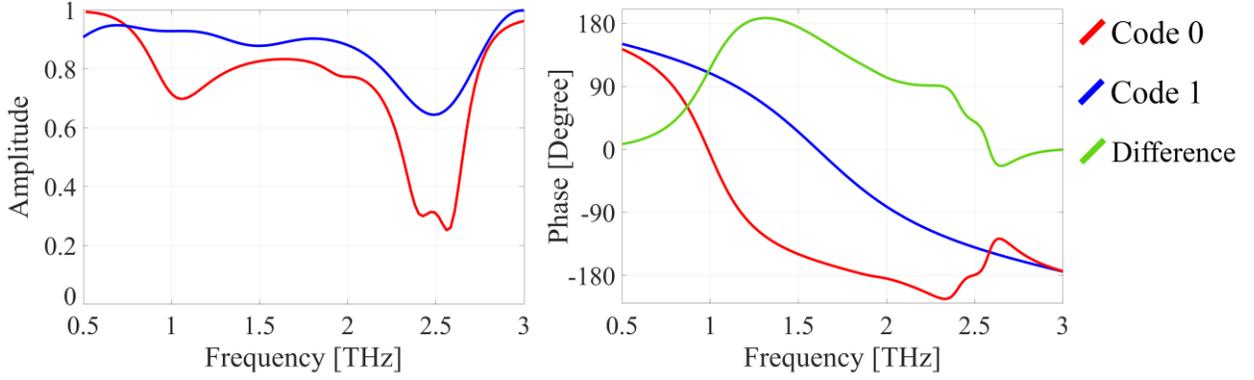

*Figure 2. The amplitude and phase spectra of the reflection coefficient of graphene-based metasurface with different chemical potentials in 0.5 THz to 3 THz bands.*

## III. PROGRAMMABLE BEAM MANIPULATION

By applying different electric field bias maps for arranged metasurface, we can switch the reflection layout in a real-time manner. The whole metasurface is occupied by 10×10 "0" and "1" tiles, while each tile contains 25 elements. The various configuration of digital particles leads to diverse functionalities only achieved by one planar formation. Here, three sample digital sequences are considered and explained. A normal y-polarized plane wave propagating along the z-direction illuminates all the elements, and the corresponding far-field patterns are depicted in Figure 3 at 1.5 THz. All the 3D far-field patterns are normalized to the maximum value of the far-field pattern belonging to the uniform coding sequence, which occurs in the boresight direction. We start with the simplest possible coding sequence "0 0 0 ..." in which all the lattices have the same reflection phase responses. Such a uniform digital sequence creates an intense reflection beam toward the normal direction. A scattering peak at the boresight direction is inevitable for the uniform digital distribution throughout the desired frequency. This layout is mimicking similar to electric and magnetic conductors. As a first presented sample distribution, "0" and "1" elements are regularly arranged in alternate rows or columns, providing the coding patterns of "0 1 0 1 ... / 0 1 0 1 ...". The combination of anti-phase contributions leads to destructive interference, producing a null in the boresight direction. In this sample, abrupt phase changes along the x-direction cause the scattered wave to split into two symmetrically oriented reflected beams. The angular direction of anomalous scattered beams can be theoretically calculated based on the generalized Snell's law of reflection [1]

$$\theta = \pm \sin^{-1}\left(\frac{\lambda_0}{2\pi}\frac{\Delta\varphi}{D}\right) \tag{4}$$

A phase difference of $\Delta\varphi=\pi$ between adjacent lattices is considered here; also D is one complete period length. Two main reflected beams at the angles of θ=29.5°, Φ=0° and θ=29.5°, Φ=180°, are calculated in Figure 3a, which has a good agreement with theoretical predictions θ=30°, Φ=0°, and θ=30°, Φ=180°. According to Equation 4, the scattered beams come closer to the boresight direction as the frequency increases, providing a beam steering functionality.

In the second example, we have investigated the proposed structure to manipulate waves in a checkerboard configuration. As we mentioned before, we need to change DC voltage distribution without any change in structure geometry. It is evident in Figure 3b that the checkerboard configuration of digital lattices with the coding sequence of "0 1 0 1 ... / 1 0 1 0 ..." makes the reflective metasurface to redirect the incident wave to four beams. By using the antenna array theory, we can predict the reflection angle by utilizing simple calculations [36]

$$\Phi = \tan^{-1}\left(\frac{\Delta\varphi_x \, D_x}{\Delta\varphi_y \, D_y}\right) \tag{5}$$

$$\theta = \sin^{-1}\left\{\frac{\lambda_0}{2\pi}\sqrt{\left(\frac{\Delta\varphi_x}{D_x}\right)^2 + \left(\frac{\Delta\varphi_y}{D_y}\right)^2}\right\} \tag{6}$$

Here, $\Delta\varphi_x$ and $\Delta\varphi_y$ are the phase differences of lattices in the x and y directions, respectively. As we mentioned before, due to structure symmetry, the reflected angle for these four redirected beams is equal, and they have a 90° difference in $\Phi$ value. The simulation reflection beam directions of $\theta$=44.1°, $\Phi$=45°, 135°, 225°, and 315° are in excellent agreement with the theoretical prediction as $\theta$=45°, $\Phi$=45°, 135°, 225°, and 315°. In the final example, we broke full symmetry in a checkerboard arrangement. In the asymmetric checkerboard configuration we employed "0 1 0 1 …/ 0 1 0 1 …/ 1 0 1 0 … / 1 0 1 0 …" encoded phase map to redirect wave in four direction. In this configuration, reflected angle $\theta$ is equal for all four reflected beams, whereas the difference between $\Phi$ angles is not equal. We can use a similar method to symmetric checkerboard and calculate the expected value for reflection angles. In this example, the simulation redirected beam directions of $\theta$=32.8°, $\Phi$=25.5°, 154.5°, 205.5°, and 334.5° are in excellent agreement with the theoretical calculation $\theta$=33.9°, $\Phi$=26.5°, 153.5°, 206.5°, and 333.5°.

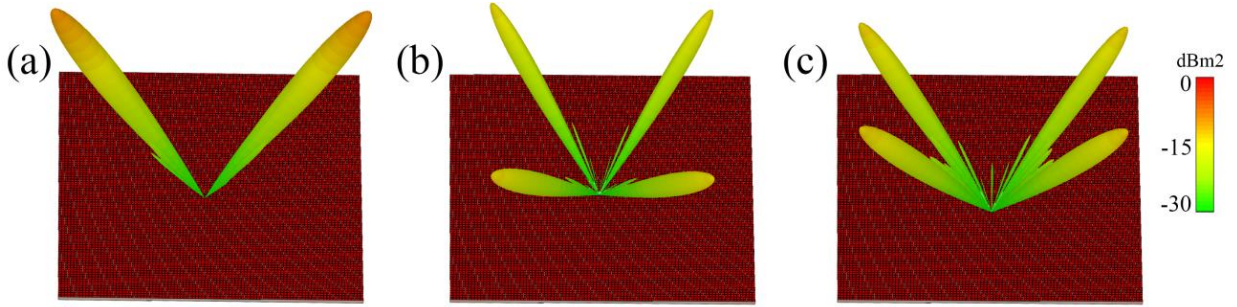

*Figure 3. The far-field simulation results of graphene-based metasurface in a) striped, b) symmetric checkerboard c) asymmetric checkerboard layout. The simulation results obtained under y-polarized illumination at 1.5 THz.*

According to introduced equations, find out the reflection angle is a function of operating frequency. So we have two influential factors to control reflected beam angles. Firstly, it is possible to control the redirected beam direction via changing illumination frequency similar to traditional works. The metasurface structure has a wide-band response and has not limited to a single frequency, so this approach is an accessible way to manipulate waves dynamically. Besides, the graphene tunability feature provides more degree of freedom to control metasurface electromagnetic response. By changing the desired encoded phase map that is equal to changing the external electric field via DC bias voltage between the layouts mentioned above or other arbitrary layouts, the reflected wave can be controlled in a real-time manner. It should be noted that all the functionalities listed are the most favorable examples of programmable coding metasurface capabilities, and real-time toggling the operational status of several lattices can attain diverse scattering behaviors. In the next section, we will investigate the structure's capability to diffuse redirected energy and absorb it to reduce reflected power in the whole half-space.

## IV. Planar Wide-Band Low-Scattering Performance

To reject the strong side lobes displayed in previous periodic configurations, irregular layouts are conventionally utilized as one of the most significant abilities of THz coding metasurfaces. We are going to explore the best random distribution whereby achieve a quasi-isotropic scattering in all directions. In this paper, a 2D-IFFT is employed to find the optimal diffusion design of the metasurface. We have introduced this approach in the "Theoretical Calculation" section comprehensively. The far-field diffused pattern of the obtained optimized arrangement at various frequencies has been shown in Figure 4. To investigate the broadband performance of the low-scattering metasurface, the far-field scattering patterns of diffusion metasurface at 1 THz, 1.5 THz, 2 THz, and 2.5 THz, are depicted in this figure.

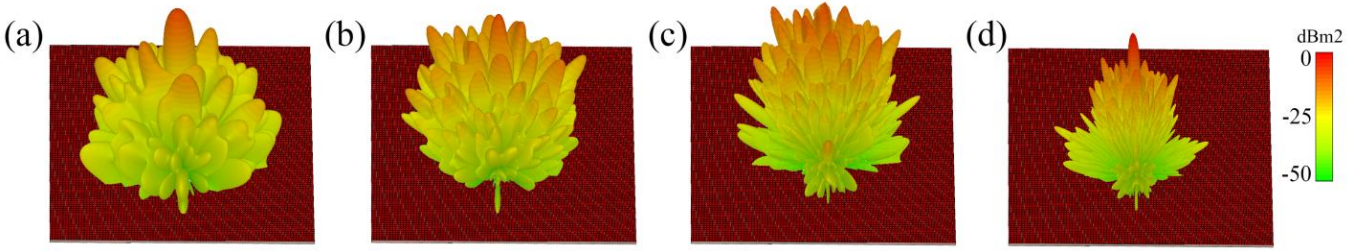

*Figure 4. The far-field results of graphene-based metasurface in the optimized distribution of graphene-based coding elements. The obtained results extracted under y-polarized illumination at a) 1 THz, b) 1.5 THz, c) 2, and d) 2.5 THz.*

As can be observed, the metasurface is capable of quasi-isotropic scattering caused by incoming waves deflected into numerous reflection beams in the first frequency band. However, in the second frequency band, the magnitude of the reflected beam reduced drastically. According to this characteristic, the reflected wave power will decrease in this band. The selected bands are adjacent together so that we can obtain a wideband frequency region for wave suppression. Because of the uncertainty in the polarization of incident waves in most practical applications, the proposed structure should provide a polarization-insensitive response under different polarization illumination. Hence, the reflection spectra of the metasurface compared with a metallic reference plate of the same size are simulated under the normal incidence with both TE and TM polarizations. The reflection spectra are obtained by comparing the scattering results in the normal direction, and the results are demonstrated in Figure 5. A low scattering feature with reflectivity reduction values less than −10 dB has been attained for both TE and TM polarizations in a broad frequency range from 1.02 to 2.82. This value is about 93.75% relative to the center frequency. The maximum specular reflection reduction is as high as 30 dB at 2.22 THz, which represent a high scattering reduction in this frequency. A reasonable similarity between TE and TM reflection results illustrates the polarization-insensitive performance of the structure at normal illumination.

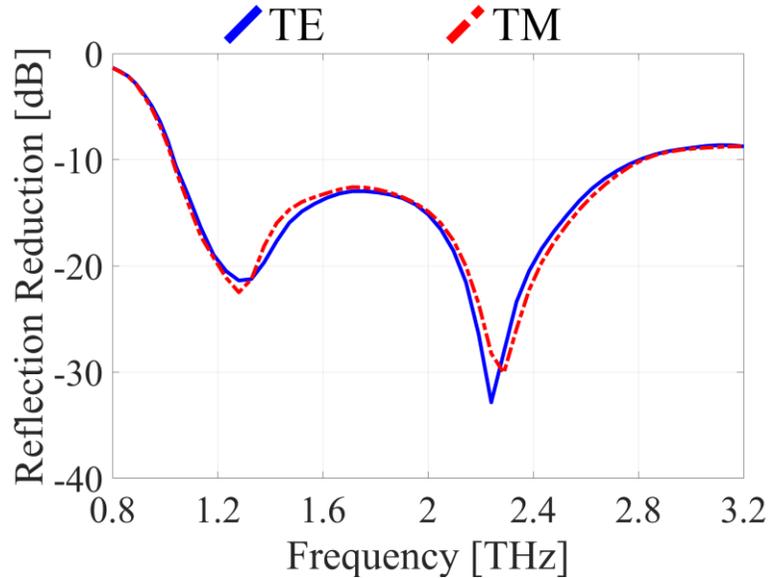

*Figure 5. The reflection reduction spectra of the optimized diffusion coding metasurface by subtracting the backscattering compared with a flat metallic plate of the same size under the TE and TM incident waves. The obtained results are related to the normal direction.*

We have study structure competence to reduce reflected wave in normal direction by utilizing random coding distribution and absorption, while the other distributions, which introduced in the previous section, has an excellent performance in normal reflection reduction. The main advantage of the optimized random arrangement is reflection reduction in all directions, not in a particular direction. For instance, in a checkerboard layout, illuminated waves reflect into four directions; as a result, reflected power in normal direction was insignificant, whereas four high power beam divided into four redirected beam. However, in the optimized layout, reflected power scattered in numerous directions and power reduction achieved in the whole half-space. To compare the diffusion performance of the optimally coding diffusion metasurfaces, quantitatively, a figure of merit is defined as:

$$\text{FOM(f)} = \frac{\text{Max [bistatic reflection of diffusion metasurface]}}{\text{Max[bistatic reflection of the metallic reference plate]}} \qquad (7)$$

Based on the energy conversion principle, by diverging the incoming waves to numerous directions, the amount of FOM has a significant reduction compared with the metallic reference plate of the same size. Therefore, As the reflected wave is diffused more uniformly toward the half-space, the bistatic scattering pattern has a smaller maximum value. In order to express the structure capability, the FOM values extracted from the corresponding result by using the mentioned equation. A metallic plate with the same size to metasurface is utilized as a comparison reference, and obtained results have been shown in Figure 6. The power of the reflected beams is strongly suppressed by more than 10 dB compared with the metallic reference plate in the desired frequency band.

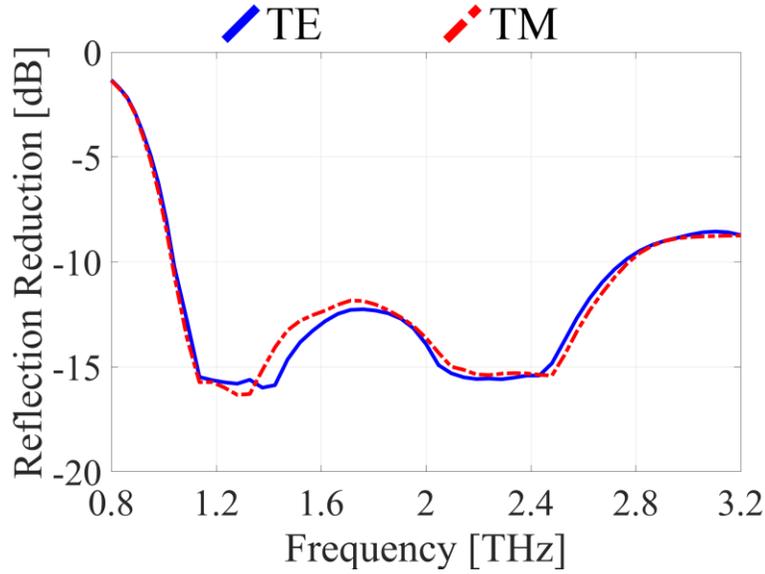

*Figure 6. Comparison of reflection reduction spectra of the optimized diffusion coding metasurface under the TE and TM incident wave. The obtained results obtained for the whole half-space based on FOM definition.*

## V. LOW-SCATTERING CYLINDRICAL STRUCTURE

As one of the various exciting characteristics of graphene, its outstanding tensile strength enables a category of graphene-based devices to be stretchable, flexible, and foldable. In more realistic scenarios, flexible and conformal reflectionless screens are also required to be capable of enclosing non-planar metallic objects [37], [38]. After showing that a random distribution of anti-phase graphene-based coding elements may achieve a planar wave diffusion, we extend this concept to a curved geometry by wrapping the metasurface around a bare metallic cylinder. Recent developments in current technologies allow us to envision such a flexible structure. The cylinder main axis is along the y-direction, and a normal incidence illuminates the whole curved metasurface in the -z-direction. The three-dimensional far-field patterns of both x- and y-polarized incident waves are given in Figure 7 at two different frequencies, 1.5 THz and 2 THz. Also, the three-dimensional far-field patterns for bare metallic cylinder has been shown in this figure for better visual comparison. In order to compare the quantitive performance of random distribution coding metasurface the reflection reduction spectra in the described case are demonstrated in Figure 7.

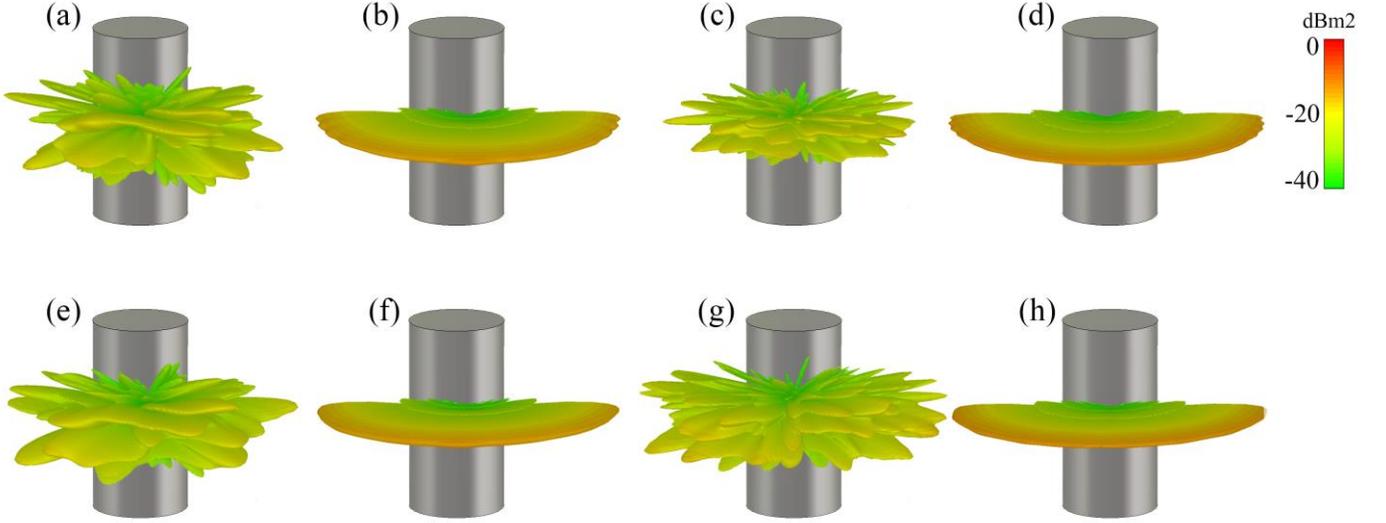

*Figure 7. Comparison between the far-field scattering patterns of the a,c,e,g) graphene-coated and b,d,f,h) bare metallic cylinder under a-d) x-polarized and e-h) y-polarized normal incidences at a,b,e,f) 1.5 THz and c,d,g,h) 2 THz.*

## VI. THEORETICAL CALCULATION

### A. Calculate Optimized Layout for Wave Diffusion

The proposed coding metasurface is capable of reflecting incoming waves to anomalous directions governed by the generalized Snell's law. The considered surface is occupied by 10×10 anti-phase super-cell. To have broadband specifications, the dimensions of each lattice are chosen between 0.6 $\lambda_{min}$ and 3 $\lambda_{max}$ in which $\lambda_{min}$ and $\lambda_{max}$ correspond to the upper and lower limits of the frequency band, respectively. Each lattice contains 5×5 basic "0" and "1" elements so it is composed of enough digital elements to minimize the mutual coupling between the neighboring coding elements. Also, each super-cell exhibits its reflection phase independently, and due to the tunable characteristic of the graphene layer, its operational status can be dynamically changed between the two states of "0" and "1". The various digital layouts of particles would result in several types of scattering patterns. According to the connection between different digital designs of the metasurface and their far-field patterns, metasurface behaviors can be analytically predicted under normal incidence. Based on the Huygens' principle, each element can be considered as a secondary source. Therefore after illumination, each coding metasurface is equivalent to an independent point source with a predetermined phase and amplitude. The far-field scattering pattern can be denoted by

$$E(\theta, \Phi) = f_{m,n}(\theta, \Phi) \sum_{m=1}^{M} \sum_{n=1}^{N} A_{m,n} e^{jmk_0 D_x \sin\theta\cos\Phi} \times e^{jnk_0 D_y \sin\theta\sin\Phi} \tag{8}$$

Where $A_{m,n} = a_{m,n}\exp(j\varphi_{m,n})$ is the complex reflection coefficient, $f_{m,n}(\theta, \Phi)$ is the pattern function of each super-cell, $k_0$ is the propagation constant, $D_x$ and $D_y$ denote super-cell dimensions, and $\theta$ and $\Phi$ are elevation and azimuth angles. Due to sub-wavelength dimensions and isotropic radiation of each super-cell, we can safely ignore super-cell pattern function during the calculation of the far-field pattern. This equation of the well-known antenna array theory is helpful for the calculation of scattering patterns caused by various digital sequences. However, by significantly increasing the number of super-cell in the structure, the computational burden arising from the slow convergence rate in Equation 8, since there are $2^{M*N}$ possible situations for the digital element arrangement. So we should use another alternative for fast optimization and calculation. In order to solve this disadvantage, a 2D inverse discrete Fourier transform is utilized to make the calculations faster. By considering the (u,v) coordinates, Equation 8 can be re-write as follows

$$E(u,v) = \sum_{m=1}^{M} \sum_{n=1}^{N} A_{m,n} e^{jmk_0 D_x u} \times e^{jnk_0 D_y v} \tag{9}$$

Where u=sin(θ)×cos(Φ), v=sin(θ)×sin(Φ). By calculating $K_{fft} \times L_{fft}$ points 2D IFFT on the complex reflection coefficients of super-cell, we have

$$IFFT(k,l) = \frac{1}{K_{fft} \times L_{fft}} \sum_{m=1}^{M} \sum_{n=1}^{N} I_{m,n} e^{jm2\pi \frac{k}{K_{fft}}} \times e^{jn2\pi \frac{l}{L_{fft}}}; k \in \{0,1,\dots,K_{fft}\}, l \in \{0,1,\dots,L_{fft}\} \tag{10}$$

We can observe significant duality relation between two upper Equations 9 and 10. Therefore the far-field pattern of the metasurface with predetermined layout can be calculated by

$$E(u,v) = K_{fft} \times L_{fft} \times IFFT(k,l) \tag{11}$$

### B. Biasing Circuits

The biasing grid is a crucial part of the meta-device, which may degrade the overall performance of the meta-device drastically. We have borrowed the phased array concept based on individual biasing and controlling the progressive phase-delay of nano-antennas in an array to vary the state of the metasurface in a real-time manner. The utilized patches are connected by skinny metallic wire links inserted in between so that the biasing voltages can be feasibly applied from one side of the metasurface. To realize this, a thin layer of $Al_2O_3$, as the back-gate, is sandwiched among graphene layers and a thin doped silicon substrate. Due to the presence of $Al_2O_3$ and doped silicon layers beneath the whole metasurface, it is enough to electrically connect the silicon layer to the DC ground from one side. The charge carriers can be accumulated on the graphene surface via applying the required external DC voltage to the biasing capacitor and yielding the desired level of chemical potential. The required DC voltage bias can be applied to each graphene layer via metallic contact where gold contact touches the silicon layer, and the chromium contact is connected to the graphene patch [39]. The chemical potential of the graphene layer can be dynamically tuned via applying an external voltage $V_g$. In the suggested biasing method, an estimated relation between $\mu_c$ and $V_g$ can be calculated by a parallel plate capacitor model [40]

$$|\mu_c| = \hbar v_F \sqrt{\frac{\pi \varepsilon_0 \varepsilon_r |V_g - V_d|}{ed}} \tag{12}$$

Where $\hbar$ is the reduced Planck constant, $v_F = 1.1 \times 10^6$ m/s denote the Fermi velocity, $\varepsilon_r$ is the relative permittivity of the dielectric $Al_2O_3$ layer, $\varepsilon_0$ is vacuum permittivity, e is an elementary charge, and d is the thickness of the supporting $Al_2O_3$ layer. Also, $V_g$ and $V_d$ are required biasing voltage and Dirac point voltage, respectively.

## VII. Conclusion

In this study, a new method to design reprogrammable graphene-based low scattering metasurface has been introduced. The proposed method yields a wide-band reflection reduction in the THz frequencies. This metasurface has provided a low reflectivity of less than −10 dB in a broad frequency band ranging from 1.02 to 2.82 THz for both TE and TM polarizations. The low-scattering performance is also preserved when the metasurface conformally wrapped around a curved metallic bar. Besides, the proposed structure has a stupendous capability for dynamic anomalous wave manipulation. Furthermore, simulation results confirm the compatibility of the proposed method for considered applications. A metasurface design method based on chemical potential optimization paves the way for designing feasible reconfigurable metasurface and versatile scattering control that would provide a wide variety of applications in the THz spectrum. Consequently, a new generation of metasurfaces with a higher degree of control will be further studied in the future works to increase the capacity of the proposed method for more extensive applications by controlling amplitude and phase in a similar frequency simultaneously. The programmable meta-device is a promising candidate to meet the THz demands for multifunctional and high technological applications.

<div align="center">REFERENCES</div>